\documentclass[10pt]{article}

 \usepackage{amsmath,amsthm,amssymb}
 \usepackage{makeidx,epsfig,lscape}
 \usepackage{color,colortbl}
 \usepackage{fancyhdr}
 \usepackage{xcolor,pict2e}

 \renewcommand{\headrulewidth}{0pt}
 \renewcommand{\footrulewidth}{0.5pt}

 \definecolor{myaqua}{rgb}{0.0,0.5,0.55}
 \definecolor{lightaqua}{rgb}{0.75,0.95,0.95}

 \usepackage[colorlinks = true,
            linkcolor = myaqua,
            urlcolor  = blue,
            citecolor = myaqua]{hyperref}

\usepackage{caption}
\usepackage{floatrow}
\def\lin#1#2{\textcolor[rgb]{0.6,0.6,0.6}{\vspace*{#1mm} \hrule
   height 3 pt \vspace*{#2mm}}}
\def\bt{\begin{tabular}}
\def\et{\end{tabular}}
\def\and{\mbox{ and }}

\def\1{{\bf 1}}

 \def\sectionn#1{\refstepcounter{section}{\color{myaqua}

 \vskip 6mm

 \noindent\Large\bf\thesection. #1}

 \vskip 3mm}
 \def\subsectionn#1{\refstepcounter{subsection}{\color{myaqua}

 \vskip 5mm

 \noindent\large\bf\thesubsection. #1}

 \vskip 2mm}

 \def\boxx#1#2#3#4#5{
 {\linethickness{#4pt}\put(#1,#5){\color{myaqua}{\line(1,0){#3}}}}
 \multiput(#1,#2)(0,#4){2}{\line(1,0){#3}}
 \multiput(#1,#2)(#3,0){2}{\line(0,1){#4}}
  }

\begin{document}

 \fancyhead[L]{\hspace*{-13mm}
 \bt{l}{\bf International Journal of Astronomy \& Astrophysics, 2019, *,**}\\
 Published Online **** 2019 in SciRes.
 \href{http://www.scirp.org/journal/*****}{\color{blue}{\underline{\smash{http://www.scirp.org/journal/****}}}} \\
 \href{http://dx.doi.org/10.4236/****.2019.*****}{\color{blue}{\underline{\smash{http://dx.doi.org/10.4236/****.2019.*****}}}} \\
 \et}
 \fancyhead[R]{\includegraphics{pic1.ps}}

 $\mbox{ }$

 \vskip 12mm

{ 

{\noindent{\huge\bf\color{myaqua}
$Spitzer$ IRAC Colors of Nebulae Associated with Star-Forming Regions}}
%
\\[6mm]
{\large\bf Yoichi Itoh$^1$, Yumiko Oasa$^2$}}
\\[2mm]
{ 
 $^1$Nishi-Harima Astronomical Observatory, Center for Astronomy, 
University of Hyogo, 
407-2, Nishigaichi, Sayo, Hyogo 679-5313, Japan\\
Email: \href{mailto:yitoh@nhao.jp}{\color{blue}{\underline{\smash{yitoh@nhao.jp}}}}\\[1mm]
$^2$Faculty of Education, Saitama University, 
255 Shimo-Okubo, Sakura, Saitama, Saitama 338-0825, Japan\\
Received Jan. 27, 2019
 \\[4mm]
Copyright \copyright \ 2019 by author(s) and Scientific Research Publishing Inc. \\
This work is licensed under the Creative Commons Attribution International License (CC BY). \\
\href{http://creativecommons.org/licenses/by/4.0/}{\color{blue}{\underline{\smash{http://creativecommons.org/licenses/by/4.0/}}}}\\
 \includegraphics{pic2.ps}

\lin{5}{7}

 { 
 {\noindent{\large\bf\color{myaqua} Abstract}{\bf \\[3mm]
 \textup{
Star-forming regions are often associated with nebulosity.
In this study, we investigated infrared diffuse emission in 
$Spitzer$ IRAC images.
The infrared nebula L1527 traces outflows emanating from 
a low-mass protostar.
The nebular color is consistent with the color of a stellar photosphere with 
large extinction.
Nebulae around the HII region W5-East are bright in the infrared.
These colors are consistent with the model color of dust containing polycyclic 
aromatic hydrocarbon (PAH).
The strength of ultraviolet irradiation of the nebulae
and the small dust fraction were deduced from the infrared colors of the 
nebulae.
We found that the edges of the nebulae are irradiated by strong ultraviolet
radiation and have abundant small dust.
Dust at the surface of the molecular cloud is thought to be  
destroyed by ultraviolet radiation from an early-type star.
 }}}
 \\[4mm]
 {\noindent{\large\bf\color{myaqua} Keywords}{\bf \\[3mm]
Interstellar medium; dust; extinction; HII regions
}

 \fancyfoot[L]{{\noindent{\color{myaqua}{\bf How to cite this
 paper:}} Y. Itoh and Y. Oasa (2019)
$Spitzer$ IRAC Colors of Nebulae Associated with Star-Forming Regions.
 International Journal of Astronomy \& Astrophysics,*,***-***}}

\lin{3}{1}

\sectionn{Introduction}

{ \fontfamily{times}\selectfont
 \noindent 
Dust is one of the most fundamental solid materials in the universe.
It is ubiquitous, but dust grains have diverse sizes.
It is widely believed that small dust is abundant in the interstellar medium.
Interstellar extinction increases with decreasing wavelength in
the ultraviolet (UV) and optical wavelengths, with a characteristic bump at 2175
 \AA.
\cite{Mathis} claimed that this wavelength dependence of the
extinction is well reproduced by graphite dust with a power law
size distribution between 0.005 and 1 $\mu$m in diameter
with an exponent of $\sim -3.5$. 
On the other hand, an overabundance of large dust has been observed
in dense regions.
\cite{Nakajima} found that the Lupus 3 dark cloud shines in 
near-infrared wavelengths. They attributed the nebular emission to
scattering of starlight by dust.
The observed near-infrared color was reproduced by
an interstellar dust model with a higher number density
for a larger size regime.
\cite{Tanii} conducted polarimetric imaging observations of a
T Tauri star, UX Tau.
With a spatial resolution of 0.1" (14 AU), they detected
a strongly polarized circumstellar disk surrounding UX Tau A.
It extends to 120 AU with a
polarization degree ranging from 1.6\% to 66\%.
The observed azimuthal profile of the polarization degree was not
consistent with dust models using Rayleigh scattering or Mie scattering 
approximations.
A thin disk model with nonspherical dust having a diameter of 60 $\mu$m 
reproduced the observed azimuthal profile.

Polycyclic aromatic hydrocarbon (PAH) is a mAstronomocal Journalor component of dust.
PAH has many broad emission features in the near- and mid-infrared wavelengths.
The strength of the PAH features depends on the number of molecules,
degree of ionization, and strength of the UV radiation.
\cite{Boersma} conducted $Spitzer$ Infrared Spectrograph (IRS) observations 
of NGC 7023,
a reflection nebula irradiated by a Herbig Be star, HD 200775.
They observed a $72" \times 54"$ region northwest of the exciting star
in the slit-scan mode.
This region contains a photodissociation region (PDR) and 
a molecular cloud.
They detected PAH emission bands between 5 and 15 $\mu$m and
fitted them by model
spectra to determine the size, charge, composition, and hydrogen adjacency of 
the dust.
The boundary between the PDR and the molecular cloud showed a distinct
discontinuity in the PAH characteristics.
In the molecular cloud, small, neutral PAHs account for the emission,
whereas
stable, large, symmetric, and compact PAH cations are abundant in the PDR.
They also found enhancement of the 6.2 and 11.0 $\mu$m emission
close to the exciting star, indicating
PAH photodehydrogenation and fragmentation.

The fluxes and shapes of PAH features can be obtained using an infrared
spectrograph.
However, the two-dimensional distribution of PAHs cannot be obtained except by
slit-scan observations or the use of a three-dimensional spectrograph.
Instead, we investigate the spatial distribution of PAHs by imaging observations
.
\cite{DL07} (hereafter DL07) synthesized the infrared emission spectra 
of dust heated by starlight. 
They considered mixtures of amorphous silicate and
graphitic dust.
Small carbonaceous dust has PAH-like properties.
The PAH mass fraction,
starlight intensity, and fraction of the dust
heated by starlight were parameters in the calculation.
The results indicated that small dust has high radiation efficiency at short
wavelengths.
They presented model emissivities for $Spitzer$ Infrared Array Camera (IRAC) and
 Multiband Imaging Photometer (MIPS) photometry.
\cite{Draine} constructed the spectral energy distributions (SEDs) of 65 
galaxies
using $Spitzer$ IRAC and MIPS photometry and James Clerk Maxwell Telescope 
Submillimetre Common-User Bolometer Array photometry.
They fitted the dust model spectra of DL07 to the SEDs
and found that small dust with PAH emission is abundant in galaxies
with high metallicity.

We investigate the infrared colors of nebulae irradiated by a nearby star
using archival $Spitzer$ IRAC data.
By comparison with the model colors, we consider the properties of dust
in molecular clouds.
In this paper, the phrase "small dust" is used for dust with less than $10^{3}$
C atoms, which corresponds to a diameter of $\sim 13$ \AA~, as it is used 
in DL07.

\renewcommand{\headrulewidth}{0.5pt}
\renewcommand{\footrulewidth}{0pt}

 \pagestyle{fancy}
 \fancyfoot{}
 \fancyhead{} 
 \fancyhf{}
 \fancyhead[RO]{\leavevmode \put(-90,0){\color{myaqua}Y. Itoh, Y. Oasa} \boxx{15}{-10}{10}{50}{15} }
 \fancyhead[LE]{\leavevmode \put(0,0){\color{myaqua}Y. Itoh, Y. Oasa}  \boxx{-45}{-10}{10}{50}{15} }
 \fancyfoot[C]{\leavevmode
 \put(0,0){\color{lightaqua}\circle*{34}}
 \put(0,0){\color{myaqua}\circle{34}}
 \put(-2.5,-3){\color{myaqua}\thepage}}

 \renewcommand{\headrule}{\hbox to\headwidth{\color{myaqua}\leaders\hrule height \headrulewidth\hfill}}

\sectionn{Data}

{ \fontfamily{times}\selectfont
 \noindent
Infrared data were taken from the Spitzer Archive Center.
We used the reduced data [post-basic calibrated data (PBCD)] of 3.6 $\mu$m 
images, 
4.5 $\mu$m images, and 5.8 $\mu$m images (PI: Fazio Giovanni).
The spatial resolution of the data is about 1.8 arcseconds.
We measured the average and standard deviation of the sky region or the 
HII region adjacent to the 
nebula and then subtracted the average count from the image.
Next, we made three types of mask images.
The first mask image was designed to reject low signal-to-noise regions.
We made it from the sky-subtracted image by replacing counts more than 
10$\sigma$ above the sky count with one and counts less than 10$\sigma$ 
above the sky count to zero.
The second mask image was designed for point source rejection.
Point sources were identified on the sky-subtracted
image by the DAOFIND task in the IRAF software, and the flux of each source was
measured by aperture photometry.
A mask was created for each source,
as a large aperture mask was made for a bright
object.
The mask regions were set to zero, and the other regions were set to unity.
We also made another point source mask image using Sextractor 
with the SEGMENTATION option in the CHECKIMAGE$\_$TYPE parameter.
These three types of mask images were multiplied by the sky-subtracted image so 
that only diffuse
emission appears in the final image.

\sectionn{Results}
\label{sec:MLE}

\subsectionn{L1527}
\label{subsec:LSE}

{ \fontfamily{times}\selectfont
 \noindent
We investigate the infrared color of a nebula without PAH emission.
L1527 is a reflection nebula associated with
the low-mass protostar IRAS 04368+2557 in the Taurus molecular cloud.
It is seen in edge-on geometry with bipolar
outflows toward the east and west.
Strong UV radiation is not expected from the central star;
thus, PAH is not expected to be excited.
Figure $\ref{L1527cc}$ shows the color-color diagram of L1527.
We measured the nebular flux of 5 $\times$ 5 pixels ($3.0'' \times 3.0''$)
in each band.
Error bars in the figure indicate the average values of standard
 deviations
of the 5 $\times$ 5 pixel regions. The error may be overestimated because
uniform color is assumed in each 5 $\times$ 5 pixel region.
In the figure, the colors of dwarfs and giants are located around (0,0)
\cite{Megeath}.
The nebular colors are plotted along the interstellar
extinction vector \cite{Chapman}
with the origin at (0,0) and are around the colors of blackbodies.
Those colors are consistent with the colors of Class I and
 Class II sources \cite{Hartmann}.
The PAH model (DL07)
We investigate the infrared color of a nebula without PAH emission.
L1527 is a reflection nebula associated with
the low-mass protostar IRAS 04368+2557 in the Taurus molecular cloud.
It is seen in edge-on geometry with bipolar
outflows toward the east and west.
Strong UV radiation is not expected from the central star;
thus, PAH is not expected to be excited.
Figure $\ref{L1527cc}$ shows the color-color diagram of L1527.
We measured the nebular flux of 5 $\times$ 5 pixels ($3.0'' \times 3.0''$)
in each band.
Error bars in the figure indicate the average values of standard
 deviations
of the 5 $\times$ 5 pixel regions. The error may be overestimated because
uniform color is assumed in each 5 $\times$ 5 pixel region.
In the figure, the colors of dwarfs and giants are located around (0,0)
\cite{Megeath}.
The nebular colors are plotted along the interstellar
extinction vector \cite{Chapman}
with the origin at (0,0) and are around the colors of blackbodies.
Those colors are consistent with the colors of Class I and
 Class II sources \cite{Hartmann}.
The PAH model (DL07)
does not reproduce the color of L1527.
We conclude that the nebula L1527 is a reflection nebula and
does not contain strong PAH emission features.
We also claim that the color-color diagram distinguishes between a reflection 
nebula and an emission nebula containing PAH features.

\begin{figure}
\begin{center}
 \begin{picture}(150,220)(0,-45)
  \put(-81,-143){\includegraphics{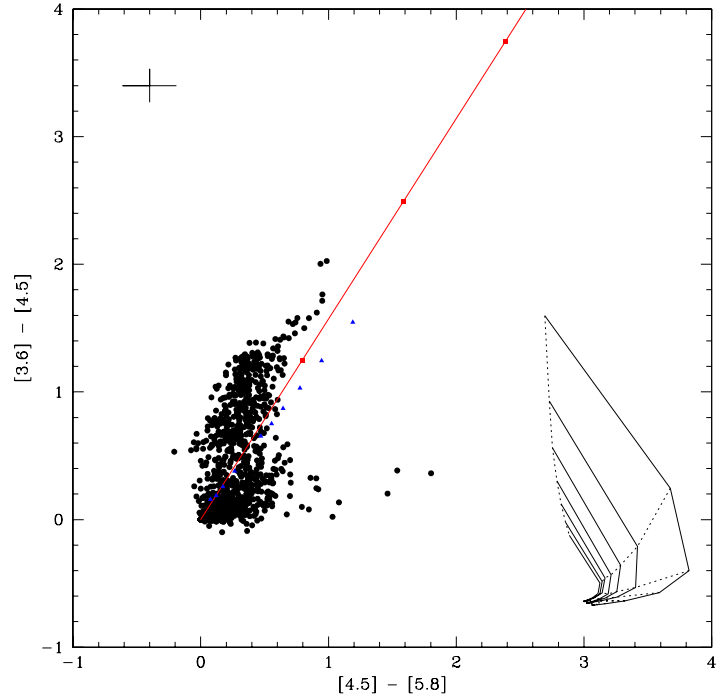}}
  \end{picture}
 \end{center}
 \caption{Infrared color of reflection nebula L1527 (filled circles).
The nebular colors are measured for every $3.0'' \times 3.0''$ region.
Stellar photospheres are located close to (0,0) for any spectral type.
A solid red line from the origin indicates the direction of int
erstellar extinction.
Filled red squares correspond to $A_{\rm V} = 100$, 200, and 30
0 mag,
respectively.
Filled blue triangles show colors of blackbodies with temperatures
between 1000 and 500 K.
The PAH colors predicted by the model are also shown at bottom right.
The color of L1527 is consistent with the colors of 
Class I and Class II sources.}
 \label{L1527cc} \end{figure}

From the figure, the amount of extinction is estimated.
The photospheric color of the protostar IRAS 04368+2557 is assumed to be (0,0), 
the same color as
dwarfs and giants.
We consider that the distance from the points of the nebular color to the
line perpendicular to the extinction vector and through the origin
corresponds to the extinction in the path from the central star
through the nebula to us.
Figure $\ref{L1527av}$ shows the extinction map of the nebula L1527.
The outflows sweep up the material in the envelope of the protostar,
forming a cavity.
The wall of the cavity or the residuals in the cavity
reflect the light from the central star.
Near the central star,
the extinction is as large as 150 mag in the $V$ band.
The extinction of the eastern outflow is slightly less than
that of the western outflow.
This estimate is consistent with a near-infrared image showing that
the eastern outflow is brighter than 
the western outflow \cite{Tamura}.

\begin{figure}
\begin{center}
 \begin{picture}(150,150)(0,-35)
  \put(-81,-88){\includegraphics{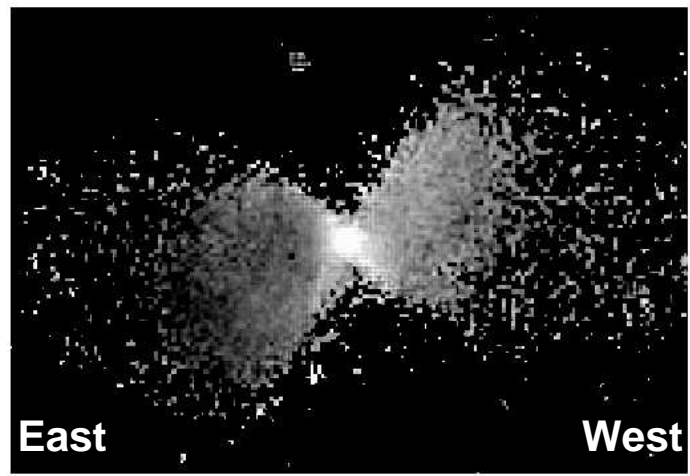}}
  \end{picture}
 \end{center}
 \caption{Extinction map of protostar in L1527. 
The field of view is 179" $\times$ 123".
A protostar is
located at the center of the image. 
$A_{\rm V} = 20$ mag and $A_{\rm V} = 150$ mag are shown in black and white, 
respectively.
Near the protostar, 
The extinction is as large as 150 mag in the $V$ band.
Outflows are located to the east (left) and west (right) of the central star 
and sweep up the material in the envelope of the protostar.
The extinction of the eastern outflow is slightly less than that of the western 
outflow.
In near-infrared wavelengths, the eastern outflow is brighter 
than the western outflow.
}
 \label{L1527av} \end{figure}

\subsectionn{W5-East}

W5 is an intensively investigated HII region.
It is located at a distance of 2.2 kpc in the Perseus arm and
constitutes a chain of molecular clouds with W3 and W4.
\cite{Koenig} searched for point sources in this region
using $Spitzer$ IRAC and MIPS images.
They identified 2064 young stellar objects on the basis of their infrared 
spectral
energy distributions.
The mAstronomocal Journalority of young stellar objects with
infrared excess belong to clusters with more than 10 members.

W5 consists of two circular H II regions, W5-East (W5-E) and W5-West.
The ionizing star in W5-E is an O7V star, BD +59$^{\circ}$ 0578.
We investigate the diffuse emission of the nebulae in the northeast region 
of W5-E.
In this region, two bright rim clouds (BRCs) have been
identified \cite{Sugitani}.
BRC 13 is classified as a Type B cloud, and BRC 14 is identified as a Type A 
cloud
according to their morphology.
It is believed that a BRC evolves from Type A through Type B 
and then Type C \cite{Miao}.
A cluster of massive young stars, AFGL 4029, is located in BRC 14.
\cite{Matsuyanagi} revealed sequential formation of low-mass
stars in this cluster.
Figure $\ref{brc13}$ is a pseudo-color map of the northeast region of W5-E.
Nebulae facing the HII region are bright in the IRAC 3 bands.
We selected several small regions to investigate
the nebular color (Figure $\ref{brc13reg}$).
Figure $\ref{brc13cc}$ shows a color-color diagram of the selected regions
of W5-E.
The colors of the nebulae are not consistent with the color of the photosphere
with extinction or with blackbody radiation.
The [4.5] $-$ [5.8] colors of the nebulae are redder than the colors of 
the photosphere and a blackbody.

\begin{figure}
\begin{center}
 \begin{picture}(150,220)(0,-35)
  \put(-81,-138){\includegraphics{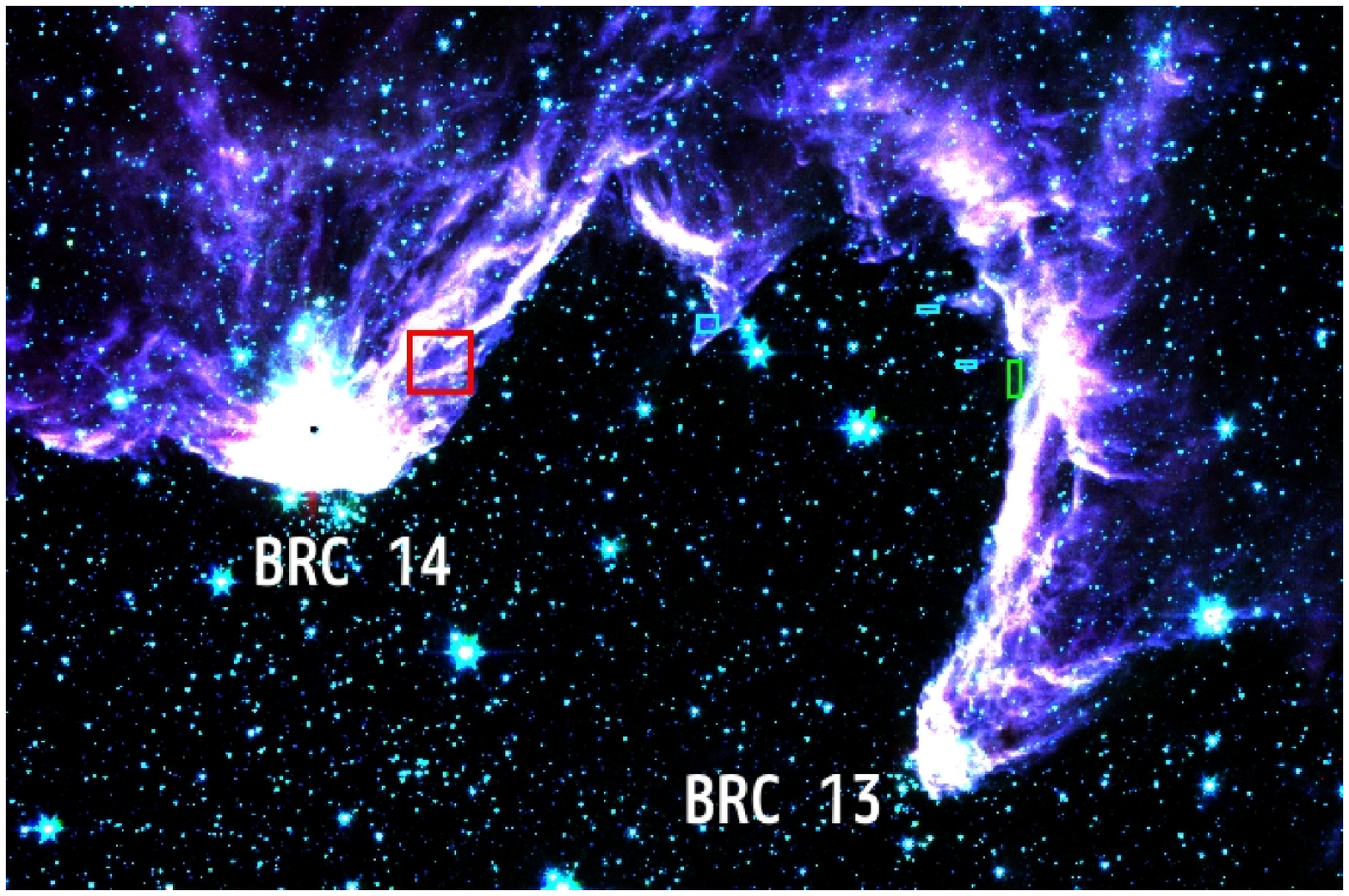}}
  \end{picture}
 \end{center}
 \caption{Pseudo-color image of northeast region of W5-E
constructed from IRAC 3.6, 4.5, and 5.8 $\mu$m images.
The field of view is 22.1' $\times$ 14.3'.
The regions whose colors are discussed on the color-color 
diagram are indicated by boxes (red: BRC 14, green: bay-like region, blue:
B- and C-like BRCs).
The exciting star is located outside the frame below the bottom of the figure.
}
 \label{brc13}
\end{figure}

\begin{figure}
\begin{center}
 \begin{picture}(150,220)(0,-35)
  \put(-81,-138){\includegraphics{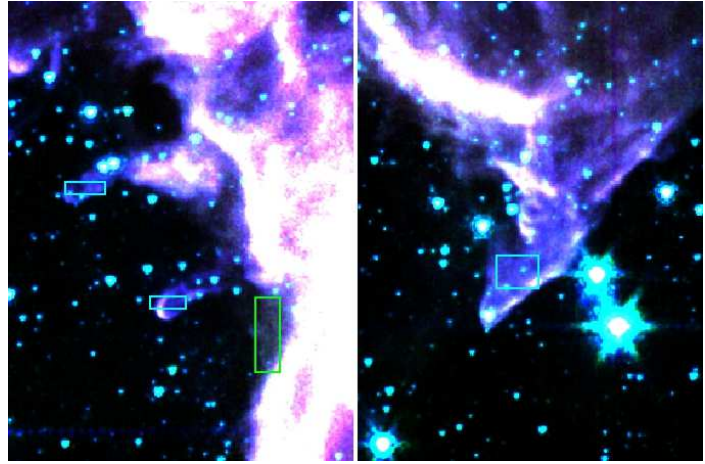}}
  \end{picture}
 \end{center}
 \caption{Close-up view of the marked regions in Figure $\ref{brc13}$. 
The field of view of both images is  
2.7' $\times$ 3.6'.
The bay-like region is indicated by a green box, and B- and C-like BRCs
are indicated by blue boxes.
}
 \label{brc13reg}
\end{figure}

DL07 calculated the IRAC colors of interstellar dust irradiated
by starlight.
They considered a mixture of amorphous silicate and graphitic dust.
The size distribution of the silicate and graphitic dust follows the power law
distributions presented by \cite{Weingartner}.
In this distribution, the volume ratio of silicate dust
to graphite dust is 1.44.
In addition, DL07 included
two populations of small carbonaceous dust;
both follow lognormal distributions with peaks at 4.0 and 20 \AA.
They constructed seven models of the Milky Way with various
mass fraction of small dust ($0.47 \% < q < 4.58 \%$).
Small carbonaceous dust has PAH-like properties.
Heating of dust by starlight is considered. Its strength
is parameterized by the factors of the interstellar radiation field 
($U_{\rm ISRF}$, \cite{Mathis83}).
They presented infrared spectra of dust with various mass
fractions of small dust and starlight intensities.
The vibration modes of PAH produce strong emission features at 3.3, 6.2,
7.6, 8.6, and 11.3 $\mu$m.
The intensities of these features change with the temperature of 
the dust.
A small dust grain is easily heated by absorption of a few UV photons
or a single UV photon
and then emits strong PAH features at short wavelengths.
Thus, the ratio of the PAH emission feature and the color of specific 
photometric
bands is a function of the fraction of small dust and the intensity of 
the UV radiation.

The observed colors of the nebulae are consistent with
the model colors of DL07 (Figure \ref{brc13cc}).
The nebulae are emission nebulae.
The colors at different positions in the nebulae are located at different loci
in the diagram.
The colors of the bay-like region are located at the top-left position 
representing a part of BRC 14 in the color-color diagram.
This indicates that the
small dust fraction is small and the UV radiation is strong
in the bay-like region
compared to those in BRC 14.
The colors of the B- and C-like BRCs are located below and to the left of 
the color of BRC 14
in the diagram.
This indicates that the B- and C-like BRCs 
have an abundant population of small dust irradiated by
or a single UV photon
and then emits strong PAH features at short wavelengths.
Thus, the ratio of the PAH emission feature and the color of specific 
photometric
bands is a function of the fraction of small dust and the intensity of 
the UV radiation.

The observed colors of the nebulae are consistent with
the model colors of DL07 (Figure \ref{brc13cc}).
The nebulae are emission nebulae.
The colors at different positions in the nebulae are located at different loci
in the diagram.
The colors of the bay-like region are located at the top-left position 
representing a part of BRC 14 in the color-color diagram.
This indicates that the
small dust fraction is small and the UV radiation is strong
in the bay-like region
compared to those in BRC 14.
The colors of the B- and C-like BRCs are located below and to the left of 
the color of BRC 14
in the diagram.
This indicates that the B- and C-like BRCs 
have an abundant population of small dust irradiated by
strong UV radiation.

\begin{figure}
\begin{center}
 \begin{picture}(150,230)(0,-35)
  \put(-81,-128){\includegraphics{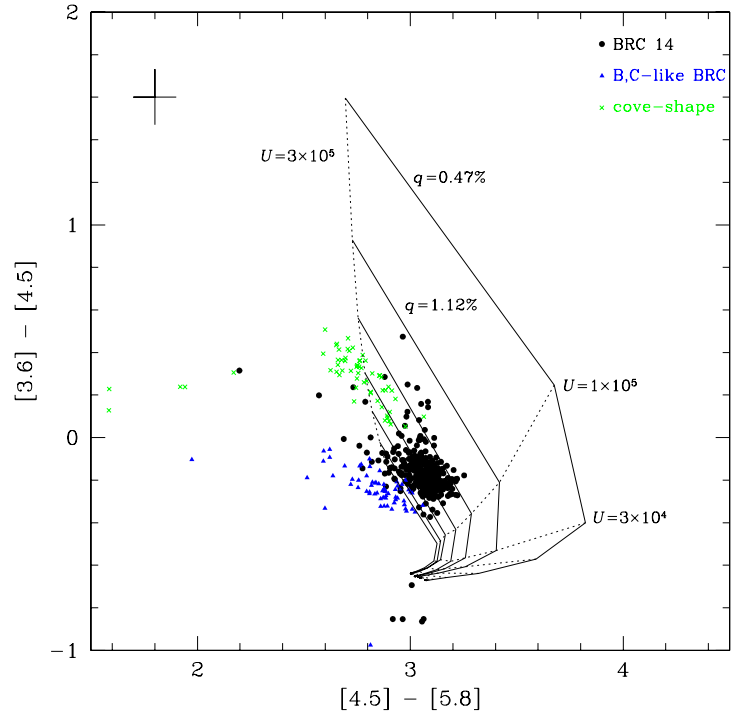}}
  \end{picture}
 \end{center}
 \caption{Infrared colors of the marked regions of the nebulae in W5-E.
Colors of different nebulae are shown by different symbols.
Overplotted are the model color of dust containing PAH.
The solid lines show the model colors of various small dust fractions 
($q$ = 0.47\%, 1.12\%, 1.77\%, 2.50\%, 3.19\%, 3.90\%, and 4.58\%).
The nebula containing abundant small dust is bluer.
The dotted lines indicate the model colors with various
UV radiation levels between $10^{2}$ times and $3 \times 10^{5}$ times the 
interstellar
UV radiation in steps of 0.5 on a logarithmic scale.
The nebulae are bright with PAH emission bands.
}
 \label{brc13cc}
\end{figure}

The strength of the UV radiation is estimated from
the nebular color.
We calculated the mean color of 5 $\times$ 5 pixel ($3.0'' \times 3.0''$).
The observed colors of the nebulae are compared
with the model color of DL07.
The model-calculated PAH fluxes for single UV strengths
are up to $3 \times 10^{5}$ times the interstellar value.
We excluded a pixel if its observed color did not match the model
color of DL07.
We also ignored pixels near the bright stars, which
are masked in the data reduction procedure.
Figure $\ref{afgl4029u}$ shows the spatial distribution of the UV strength.
As expected, the edge of the nebulae facing the HII region show
colors indicating strong UV radiation. 
At the inner part of the nebulae, the colors are 
consistent with the model color with weak UV radiation.
Most of the UV photons from the exciting star
are thought to be absorbed and scattered by dust in the molecular cloud.
We also estimate the small dust fraction as described above.
Figure $\ref{afgl4029d}$ shows the spatial distribution of the small dust 
fraction in W5-E.
The distribution is not uniform in the nebulae. 
At the inner part of the nebulae, the small dust fraction is small,
and at the edge of the nebulae, it is large.
This inhomogeneity is especially obvious for BRC 13, a Type B BRC.

The fact that small dust is more abundant at the edge of the nebulae
than in the inner part of the nebulae
indicates that large dust grains at the surface of the molecular clouds
are reduced to small dust grains
by UV radiation from the early-type star.

\begin{figure}
\begin{center}
 \begin{picture}(150,220)(0,-35)
  \put(-81,-138){\includegraphics{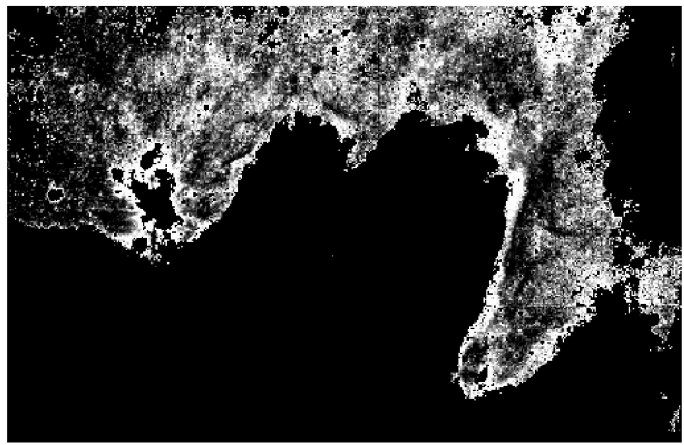}}
  \end{picture}
 \end{center}
 \caption{UV irradiation of the nebulae.
$\log (U) = 5.15$ mag and $\log (U) = 5.30$ mag are depicted in black and white,
 respectively.
The edge of the nebulae facing the HII region are irradiated
by strong UV radiation.
}
 \label{afgl4029u}
\end{figure}

\begin{figure}
\begin{center}
 \begin{picture}(150,220)(0,-35)
  \put(-81,-138){\includegraphics{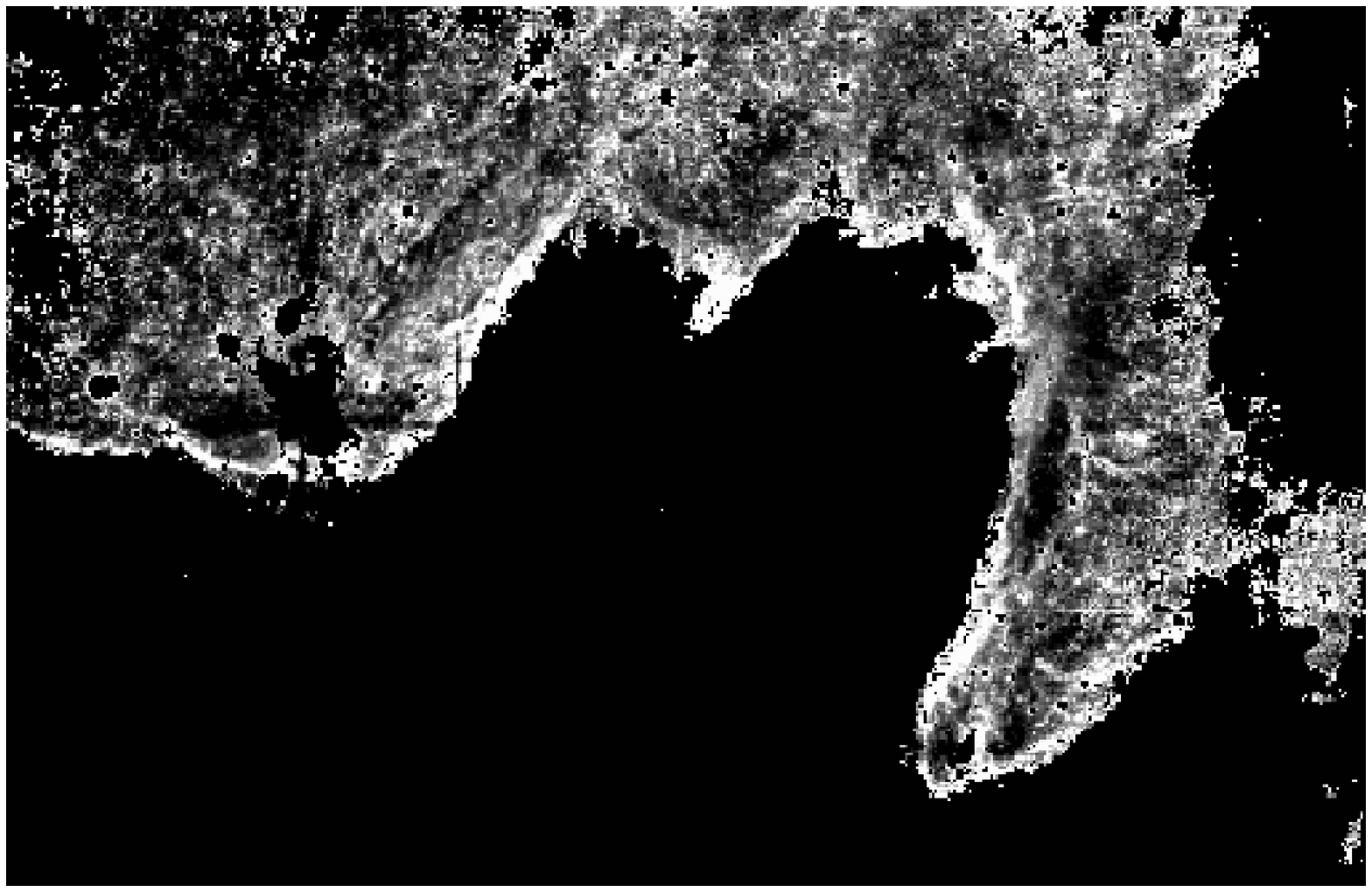}}
  \end{picture}
 \end{center}
 \caption{Spatial distribution of small dust fraction.
$q = 1.5 \%$ mag and $q = 3.0 \%$ mag are depicted in black and white, 
respectively.
The map indicates a large small dust fraction at the edge of the nebulae.
}
 \label{afgl4029d}
\end{figure}
}

\sectionn{Discussion}

{ \fontfamily{times}\selectfont
 \noindent
Our claim that dust grains at the surface of the molecular cloud
are destroyed by UV radiation emanating from an early-type star is 
deduced from the nebular color in the flux-calibrated images.
To estimate the precise flux of the 
nebulae, we subtracted the flux of the HII region adjacent to the nebulae.
We examined whether the uncertainty in the flux of the HII region challenges
our claim.
We added or subtracted a flux representing 5 times the standard deviation 
of the 
flux of the HII region and then estimated the strength of the UV radiation
and fraction of small dust.
Even after we added or subtracted the flux,
the claim that small dust is abundant at the surface of the 
molecular cloud remained valid.
Diffuse emission appears even in the HII region (see Figure 1
of \cite{Koenig}).
Even if the UV strength and dust population are estimated using
the original PBCD images,
the spatial distribution of the dust population does not change;
i.e., small dust is abundant at the surface of the molecular cloud.

The overabundance of small dust at the surface of the molecular cloud
is deduced from the ``blue'' infrared colors at the edge of the nebula.
Nebulae have different infrared colors not only with different small dust 
fractions
and different UV radiation strengths but also with
different amounts of extinction.
The blue color at the edge of the nebula
is naturally thought to be produced by the low extinction.
At the edge of the nebulae, the small dust fraction is 
as high as 4\%, whereas at the inner part of the nebulae,
the fraction is as low as 2\% and reaches 1.5\% locally.
The [4.5] $-$ [5.8] color difference between the edge and the region 
with $q=1.5\%$
is about 0.2 mag (Figure 5).
These observations can be reproduced if the inner part of the nebulae
has an extinction of 30 mag in the $V$ band.
\cite{Matsuyanagi} conducted a near-infrared imaging survey of 
the AFGL 4029 cluster in BRC 14.
They detected 605 sources, most of which are classified as background stars.
The visual extinction of a star can be estimated from the color difference
between the observed color and the intrinsic color of a dwarf or a giant.
The color-color diagram of the sources detected in \cite{Matsuyanagi}
indicated that the largest visual extinction is $\sim$20 mag.
The average extinction of the sources in the AFGL 4029 cluster and 
between the cluster and the bright rim is 7.8 mag in the $V$ band.
Thus, the visual extinction does not produce the infrared color difference
between the edge and the inner part of the nebula.
Note that our line of sight and the incident
direction of UV radiation from the early-type star to the molecular cloud
are different.
Nevertheless, there is no robust evidence that the observed 
color difference of the infrared nebula is totally reproduced
by the difference in the extinction.
We think that the observed color difference is caused, at least partially, 
by the spatial inhomogeneity of the small dust fraction.

,

The process of dust destruction has been intensively investigated.
\cite{Salpeter} claimed that thermal evaporation of dust due to strong
radiation from an exciting star is likely to be unimportant even
in an HII region.
Similar arguments were given in \cite{Draine79}.
On the other hand, recent observational studies indicated spatial
variation of the physical or chemical state of PAH.
\cite{Verstraete} observed the M17-Southwest
photodissociation region with the $Infrared Space Observatory$ Short Wavelength 
Spectrometer.
They found that all the PAH features peak at the boundary
of the HII region and the molecular cloud,
indicating spatial variation of the PAH abundance.
From the flux ratios of the PAH features, they concluded that
most of the PAHs are ionized and almost fully hydrogenated
in the HII region and at the boundary.
\cite{Pavlyuchenkov} constructed a radiative transfer
model of ionized hydrogen regions and fitted it to the 
infrared spectra of the HII region RCW 120.
From the spatial variation of the PAH emission features,
they concluded that PAHs are destroyed in the ionized region
on a timescale of 30 Myr.
Observational study of spatially resolved dust size distributions
in molecular clouds under different UV environments will conclusively 
identify the dust destruction process.
}

\sectionn{Conclusions}

{ \fontfamily{times}\selectfont
 \noindent
We investigated the infrared colors of nebulae associated with star-forming 
regions using $Spitzer$ IRAC images.
\begin{itemize}
\item The color of the diffuse emission of L1527 is consistent with
a photospheric color with large extinction.
The central source of the emission is a low-mass protostar, which does 
not emit strong UV radiation that excites PAHs.
\item The nebulae in W5-E facing the HII region are bright in the infrared.
Their color is not uniform.
A comparison with the model color of small dust emission
revealed that the nebulae are bright in PAH emission features.
The nebular color indicated that the nebulae facing the HII region
are irradiated by strong UV radiation and contain a large population
of small dust.
We claim that the dust grains at the surface of the molecular cloud
are destroyed by UV radiation from an early-type star.
\end{itemize}

 {\color{myaqua}

 \vskip 6mm

 \noindent\Large\bf Acknowledgments}

 \vskip 3mm

{ \fontfamily{times}\selectfont
 \noindent
 We thank the editor and the referee for their comments.
This work is based on observations made with the Spitzer Space Telescope,
which was operated by the Jet Propulsion Laboratory, California Institute
of Technology under a contract with NASA.
This work was supported by JSPS KAKENHI Grant Number JP24540231.

 {\color{myaqua}

}}

\end{document}